\documentclass[preprintnumbers,a4paper,twocolumn,floats,nofootinbib,superscriptaddress,aps,prd]{revtex4-1}
\usepackage{graphicx,gensymb}
\usepackage{amsmath,amssymb,amsfonts,mathtools}
\usepackage{hyperref}
\usepackage{slashed,wasysym}

\newcommand{\sln}{\sin l_{\rm{CP}}} 
\newcommand{\cln}{\cos l_{\rm{CP}}}
\newcommand{\san}{\sin\alpha_{\rm{GP}}} 
\newcommand{\can}{\cos\alpha_{\rm{GP}}} 
\newcommand{\sdn}{\sin\delta_{\rm{GP}}} 
\newcommand{\cdn}{\cos\delta_{\rm{GP}}} 

\newcommand{\cze}{\cos\zeta_{\rm{A}}} 
\newcommand{\sze}{\sin\zeta_{\rm{A}}} 
\newcommand{\cz}{\cos z_{\rm{A}}} 
\newcommand{\sz}{\sin z_{\rm{A}}} 
\newcommand{\ct}{\cos \theta_{\rm{A}}} 
\newcommand{\st}{\sin \theta_{\rm{A}}} 

\newcommand\fdg{\hbox{$.\!\!^\circ$}}
\newcommand\farcm{\hbox{$.\!\!^{\prime}$}}
\newcommand\farcs{\hbox{$.\!\!^{\prime\prime}$}}

\newcommand{\Er}{E_{\rm{R}}}

\SetSymbolFont{symbols}{bold}{OMS}{cmsy}{b}{n}
  \DeclareSymbolFont{bmisymbols}{OML}{cmm}{b}{it}
  \DeclareMathSymbol{\bepsilon}{0}{bmisymbols}{"0F}

\newcommand{\epx}{\hat{\bepsilon}_{\mathbf{x}}}
\newcommand{\epy}{\hat{\bepsilon}_{\mathbf{y}}}

\newcommand{\epone}{\hat{\bepsilon}_{\mathbf{1}}}
\newcommand{\eptwo}{\hat{\bepsilon}_{\mathbf{2}}}

\begin{document}
\title{The Earth's velocity for direct detection experiments}

\preprint{IPPP/13/95, DCPT/13/190}

\author{Christopher McCabe}
\affiliation{Institute for Particle Physics Phenomenology, Durham University, South Road, Durham, DH1 3LE, United Kingdom\\
{ \tt  \footnotesize \href{mailto:christopher.mccabe@durham.ac.uk}{christopher.mccabe@durham.ac.uk}} \smallskip}

\begin{abstract} 
The Earth's velocity relative to the Sun in galactic coordinates is required in the rate calculation for direct detection experiments. We provide a rigorous derivation of this quantity to first order in the eccentricity of the Earth's orbit. We also discuss the effect of the precession of the equinoxes, which has hitherto received little explicit discussion. Comparing with other expressions in the literature, we confirm that the expression of Lee, Lisanti and Safdi is correct, while the expression of Lewin and Smith, the {\it de facto} standard expression, contains an error. For calculations of the absolute event rate, the leading order expression is sufficient while for modulation searches, an expression with the eccentricity is required for accurate predictions of the modulation phase. \end{abstract} 

\maketitle

\section{Introduction}

A convincing body of evidence indicates that there is a large population of particle dark matter in our galaxy~\cite{Bovy:2012tw}. One of the most promising experimental probes of the non-gravitational interactions of dark matter is direct detection experiments. By measuring the nuclear recoil energy spectrum, they aim to reconstruct the mass and interaction cross-section for dark matter to interact with nuclei~\cite{Goodman:1984dc}. 

For dark matter with local density $\rho_{\rm{DM}}$, mass $m_{\rm{DM}}$ and differential scattering cross-section $d\sigma/d\Er$ to interact with a nucleus of mass $m_{\rm{N}}$, the differential scattering rate (in units of counts/kg/day/keV) is
\begin{equation}
\frac{d R}{d \Er}=\frac{\rho_{\rm{DM}}}{m_{\rm{N}} m_{\rm{DM}}}\int_{v_{\rm{min}}}d^3 v \frac{d \sigma}{d \Er} v f_{\rm{det}}(\mathbf{v},t) \;,
\end{equation}
where $v=|\mathbf{v}|$ is the dark matter speed, $f_{\rm{det}}$ the velocity distribution in the rest frame of the detector and $v_{\rm{min}}$ is the minimum speed that the dark matter must have in order that a nucleus recoils with energy $\Er$. The velocity distribution in the detector's rest frame is simply related to the distribution in the galactic rest frame $f_{\rm{gal}}$ (for which numerous models are known e.g.~\cite{Binney2008,Evans:2000gr,Evans:2005tn,Chaudhury:2010hj,Bozorgnia:2013pua,Hunter:2013vua,Fornasa:2013iaa}), through a Galilean boost
\begin{equation}
f_{\rm{det}}(\mathbf{v},t)=f_{\rm{gal}}(\mathbf{v}+\mathbf{v}_{\rm{Earth}}(t))\;.
\end{equation}
Here, $\mathbf{v}_{\rm{Earth}}(t)$ is the velocity of the Earth with respect to the galactic rest frame and it is conventionally decomposed into three terms
\begin{equation}
 \mathbf{v}_{\rm{Earth}}(t)=\mathbf{v}_{\rm{LSR}}+\mathbf{v}_{\rm{pec}}+\mathbf{u}_{\rm{E}}(t)\;,
 \end{equation}
the velocity of the local standard of rest (LSR), the peculiar velocity of the Sun with respect to the LSR and the velocity of the Earth around the Sun. These quantities are normally expressed in galactic rectangular coordinates in which $X$ points towards the galactic centre, $Y$ in the direction of galactic rotation and $Z$ normal to the galactic plane in the direction of the galactic North pole. In this case $\mathbf{v}_{\rm{LSR}}=(0,v_0,0)$, where conventionally $v_0=220$~km/s and \mbox{$\mathbf{v}_{\rm{pec}}=(11.1\pm1.2,12.2\pm2.0,7.3\pm0.6)$}~km/s~\cite{Schoenrich:2009bx}. Unfortunately there are significant uncertainties on both of these parameters with recent inferences typically finding higher values, see e.g~\cite{Schonrich:2012qz,Bovy:2012ba}. The impact on experimental exclusion limits due to the uncertainty in $v_0$ and $\mathbf{v}_{\rm{pec}}$ has been discussed previously~\cite{McCabe:2010zh}.

Given its central importance in evaluating the event rate, it is surprising that there is no rigorous derivation of $\mathbf{u}_{\rm{E}}(t)$ in the literature. The velocity $\mathbf{u}_{\rm{E}}(t)$ is especially important for annual modulation searches~\cite{Drukier:1986tm, Freese:1987wu}, which are growing in prominence with results from DAMA~\cite{Bernabei:2013xsa}, CoGeNT~\cite{Aalseth:2011wp} and CDMS~\cite{Ahmed:2012vq}, since all of the time dependence in the rate enters through $\mathbf{u}_{\rm{E}}(t)$. The {\it de facto} standard for $\mathbf{u}_{\rm{E}}(t)$ is the result of Lewin and Smith~\cite{Lewin:1995rx}, which includes the first order correction from the eccentricity of the Earth's orbit. In this paper, we provide a rigorous derivation of $\mathbf{u}_{\rm{E}}(t)$ and we demonstrate that the expression of Lewin and Smith contains an error, as suggested in~\cite{Lee:2013xxa}.

In section~\ref{section:orbit} we introduce our notation and define the relevant parameters that describe the Earth's orbit around the Sun. We give two equivalent expressions for $\mathbf{u}_{\rm{E}}(t)$ in section~\ref{section:Eu}. Both of these expressions are series expansions to first order in the eccentricity of the Earth's orbit and in the epoch of date, a parameter that accounts for the precession of the equinoxes. The effect of this precession on $\mathbf{u}_{\rm{E}}(t)$ has not previously been discussed in the literature. Section~\ref{section:phenomenology} discusses the phenomenological consequences from using different expressions for $\mathbf{u}_{\rm{E}}(t)$ when calculating observable quantities at direct detection experiments. Finally, two appendices contain technical details.

\section{The Earth's orbit}
\label{section:orbit}

\begin{figure}[t!]
\includegraphics[width=0.99\columnwidth]{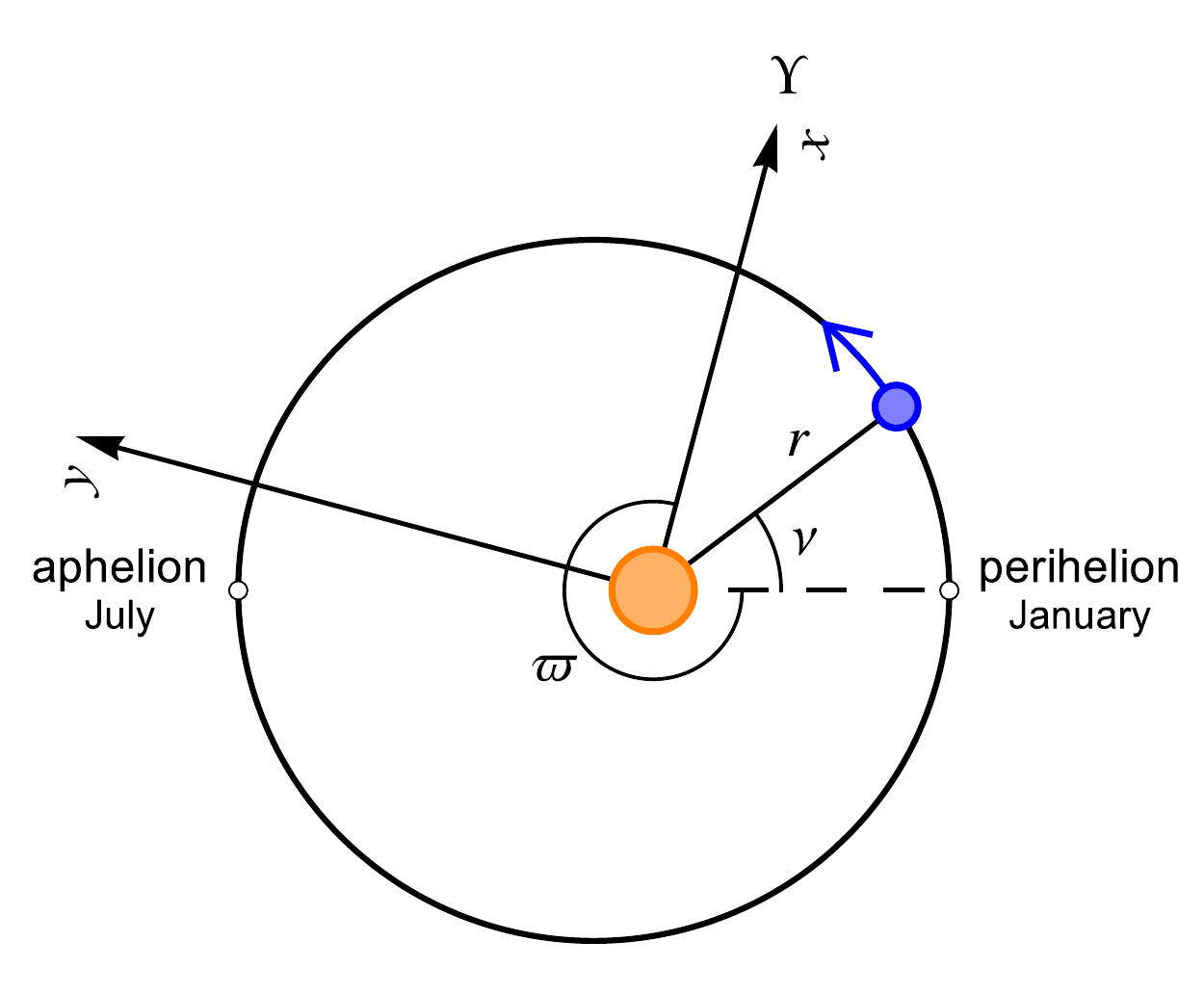}
\caption{The elliptical orbit of the Earth (blue disc) around the Sun (yellow disc). The eccentricity of this ellipse is ten times the eccentricity of the Earth's orbit ($10\times0.01671$). The direction of the orbit is anti-clockwise, as indicated by the blue arrow.  The Earth-Sun radius is $r$, the true anomaly is $\nu$ and the longitude of perihelion is $\varpi$. The $x$ and $y$ axes of the heliocentric ecliptic rectangular coordinate system are shown. The $x$-axis points in the direction of the vernal equinox ($\Upsilon$) and the $y$-axis points in the direction of the summer solstice. Ecliptic longitude $\ell=\varpi+\nu$ is measured anti-clockwise from the $x$-axis.}
\label{fig:ellipse}
\end{figure}

The Earth's orbit around the Sun is described by Kepler's laws. The purpose of this section is to introduce the parameters that are relevant for our discussion while leaving a more comprehensive discussion to standard texts on the subject e.g.~\cite{Murray1999}.

As shown in fig.~\ref{fig:ellipse}, the Earth has an elliptical orbit around the Sun. With the Sun at one of the focus points of the ellipse, the Earth-Sun radius is
\begin{equation}
r=\frac{a (1-e^2)}{1+e \cos\nu}\;,
\end{equation}
where $a=1$~AU is the length of the orbit's semi-major axis, $e=0.01671$ is the eccentricity and $\nu$ is the true anomaly, the angle between the Sun-Earth vector and the direction of the perihelion, as shown in fig.~\ref{fig:ellipse}. The angle $\nu$ does not increase uniformly with time, therefore it is conventional to relate it to the mean anomaly $g$, which does ($g$ does not have a direct geometric interpretation but it is related to the eccentric anomaly, which does). There is no closed form expression relating $\nu$ and $g$ but owing to the smallness of $e$, a power series in $e$ is a good approximation~\cite{Murray1999}
\begin{equation}
\label{eq:centre}
\nu=g+2e \sin g+\frac{5}{4}e^2 \sin 2g+\mathcal{O}(e^3)\;.
\end{equation}
This power series is often call `the equation of the centre'.

The numerical values that we quote for these parameters must be with respect to a spatial coordinate system and a reference time. The $x$ and $y$ axes of our right-handed spatial coordinate system are shown in fig.~\ref{fig:ellipse}. We will refer to this as the  heliocentric ecliptic rectangular coordinate system since the $x-y$ axes span the ecliptic plane and the Sun is at the origin. The $x$-axis points in the direction of the Earth when it is at the vernal equinox $\Upsilon$, the $y$-axis is orthogonal pointing in the direction of the Earth when it is at the (northern hemisphere) summer solstice and the $z$-axis points out of the page. This convention differs from some others in the literature in which the $x$-axis points towards the autumnal equinox and the $y$-axis towards the winter solstice. The reference time from which other times will be referred is the J2000.0 epoch, which refers to 12h (Terrestrial Time) on 1st January 2000. To an excellent approximation, this is the same as noon GMT on 1st January 2000.

The ecliptic longitude $\ell$ and latitude $b$ measure the angular distance from the $x$-axis and the distance from the $x-y$ plane. The rectangular coordinates of a point on the unit sphere are related to $\ell$ and $b$ through $(x,y,z)=(\cos \ell \cos b, \sin \ell \cos b, \sin b)$. To an excellent approximation, the Earth's orbit remains in the ecliptic plane so $b=0$, while from fig.~\ref{fig:ellipse} we see that the ecliptic longitude is related to $\nu$ and $\varpi$ through
\begin{align}
\ell&=\varpi+\nu\\
&= L+2e \sin g+\frac{5}{4}e^2 \sin 2g+\mathcal{O}(e^3)\;,
\end{align}
where $L=\varpi+g$ is the mean longitude, $\varpi$ is the longitude of perihelion and the eccentricity, expressed in degrees is $e=0\fdg9574$.

Following~\cite{Lewin:1995rx}, for this work we take parameter values from~\cite{AA2014}. They quote 
\begin{align}
L&=280\fdg460+0\fdg9856474 \,n\\
g&=357\fdg528+0\fdg9856003\,n
\end{align} 
so that 
\begin{equation}
\varpi=282\fdg932+0\fdg0000471\,n\;.
\end{equation}
Here $n$ is the (fractional) day number from J2000.0. Appendix~\ref{sec:epoch} describes how to calculate $n$ given the calendar date. These parameter values are accurate to $1\farcm0$ between 1950 and 2050~\cite{AA2014}.

\section{The Earth's velocity around the Sun}
\label{section:Eu}

We now derive two forms for $\mathbf{u}_{\rm{E}}(t)$, the Earth's velocity relative the Sun in galactic coordinates. Although these forms are superficially different, we show that they are equivalent. The first form results from projecting the Earth's position onto unit vectors that are parallel to the $x$ and $y$ axes of the heliocentric ecliptic coordinate system of fig.~\ref{fig:ellipse}. The position and velocity in galactic coordinates follows when these unit vectors are expressed in galactic coordinates. The second form results from projecting the Earth's position onto unit vectors parallel to the axes of the galactic coordinate system, whose position is given in terms of $\ell$ and $b$. The components of the Earth's velocity in galactic coordinates in terms of $\ell$ and $b$ follows straightforwardly.

\subsection{Projection onto ecliptic axes}
\label{subsection:ecliptic}

The position of the Earth relative to the Sun, expressed in terms of unit vectors $\epx$ and $\epy$ that point in the direction of the $x$ and $y$-axes of the heliocentric ecliptic coordinate system shown in fig.~\ref{fig:ellipse} is
\begin{equation}
\mathbf{r}=r \cos\ell\, \epx+r \sin \ell\, \epy\;.
\label{eq:r-ecl}
\end{equation}
The velocity as a power series follows by differentiating this expression and expanding $r$ and $\ell$ in terms of $e$. The series can in principle be expanded to any order but in practice, an expansion to first order in $e$ is sufficient. In that case, the velocity is
\begin{equation}
\label{eq:uex1}
\begin{split}
\mathbf{u}_{\rm{E}}(t)&=-\langle u_{\rm{E}}\rangle\left(\sin L+e\sin(2 L-\varpi)\right)\epx\\
&\quad+\langle u_{\rm{E}}\rangle\left(\cos L+e\cos(2 L-\varpi)\right)\epy
\end{split}
\end{equation}
with $\langle u_{\rm{E}}\rangle=a\dot{L}=29.79$~km/s. Here we have neglected time derivatives of $\epx$ and $\epy$ $(|\dot{\hat{\bepsilon}}_{\mathbf{i}}/\hat{\bepsilon}_{\mathbf{i}}\lesssim10^{-5}|)$ and ignored terms of order $\dot{\varpi}/\dot{L}\approx5\times10^{-5}$. 

The problem is now reduced to finding expressions for $\epx$ and $\epy$ in galactic coordinates. Let $-\mathcal{R}$ and $\mathcal{M}$ be the transformation matrices from heliocentric ecliptic rectangular coordinates to equatorial rectangular coordinates and from equatorial rectangular coordinates at J2000.0 to galactic rectangular coordinates respectively. Then, $\epx=-\mathcal{M}\mathcal{P}^{-1}\mathcal{R}\,\mathbf{x}_{\text{hel-ecl}}$ and $\epy=-\mathcal{M}\mathcal{P}^{-1}\mathcal{R}\,\mathbf{y}_{\text{hel-ecl}}$ with $\mathbf{x}_{\text{hel-ecl}}=(1,0,0)$ and $\mathbf{y}_{\text{hel-ecl}}=(0,1,0)$. The elements of these rotation matrices are given in appendix~\ref{app:rotation}.

These expressions also include the inverse of the matrix $\mathcal{P}$. The inverse of $\mathcal{P}$ is the precession matrix that rotates equatorial rectangular coordinates at epoch of date $T$ to equatorial rectangular coordinates at J2000.0. This is required because $\mathcal{M}$ relates equatorial and galactic coordinates only at J2000.0. The physical origin of the precession (`the precession of the equinoxes') is due to the torque on the Earth from the gravitational interaction with the Moon, Sun and planets~\cite{Green:2003yh}. It causes the position of stars and in particular, the right ascension $\alpha_{\rm{GP}}$ and declination $\delta_{\rm{GP}}$ of the north Galactic pole and the longitude $l_{\rm{CP}}$ of the north celestial pole (which enter $\mathcal{M}$) to precess in the equatorial coordinate system. The same gravitational torques cause a separate effect called nutation, however we ignore it because it is always sub-dominant compared with the effect of precession~\cite{Green:2003yh}. 

Carrying out the matrix multiplication and expanding $\epx$ and $\epy$ to first order in $T$, we find that
\begin{align}
\label{eq:epx}
\epx&=
 \begin{pmatrix}
0.054876\\
-0.494109 \\
0.867666
 \end{pmatrix}+ \begin{pmatrix}
-0.024232 \\
-0.002689  \\
1.546\times10^{-6}
 \end{pmatrix}\,T\\
\epy&=
 \begin{pmatrix}
0.993821\\
 0.110992 \\
0.000352
 \end{pmatrix}+\begin{pmatrix}
0.001316\\
-0.011851 \\
0.021267
 \end{pmatrix}\,T\;.
 \label{eq:epy}
\end{align}
For practical use, the epoch of date $T$ is related to the day from J2000.0 through
\begin{equation}
T=\frac{n}{36525}\;.
\end{equation}

\subsection{Projection onto galactic axes}
\label{subsection:galactic}

For the second form, the position vector of the Earth in galactic coordinates is found by projecting onto the galactic axes $X$, $Y$ and $Z$. Expressing the direction of the $X$, $Y$ and $Z$ in terms of latitude and longitude $(b_i,\ell_i)$, where $i=X,Y,Z$, the position vector for each component is 
\begin{equation}
\label{eq:r-gal}
\mathbf{r}_i=r \cos b_i \cos(\ell-\ell_i)\;.
\end{equation}
As with eq.~\eqref{eq:r-ecl}, the velocity follows by differentiating this equation and expanding $r$ and $\ell$ in powers of $e$. To first order in $e$, the velocity for each component $i$ is
\begin{equation}
\label{eq:uex2}
\begin{split}
\mathbf{u}_{\mathrm{E}\,i}(t)&=\langle u_{\rm{E}}\rangle\cos b_i\left[ \sin(L-\lambda_i)\right.\\
&\qquad\qquad\qquad\qquad\left.+e \sin(2 L-\lambda_i-\varpi)  \right]\;,
\end{split}
\end{equation}
where, for convenience, we define $\lambda_i\equiv\ell_i+180\degree$ to absorb an overall minus sign that would otherwise be present. As with eq.~\eqref{eq:r-ecl}, $\langle u_{\rm{E}}\rangle=a\dot{L}=29.79$~km/s and we have neglected corrections of order $\dot{\varpi}/\dot{L}$, $\dot{\lambda}_i/\lambda_i$ and $\dot{b}_i/b_i$, all of which are smaller than $\sim10^{-5}$.

To find $b_i$ and $\ell_i$, we first find the unit vectors that point in the direction of the $X$, $Y$ and $Z$ axes in heliocentric ecliptic coordinates through the relation $\mathbf{x}_{\text{hel-ecl}}=-\mathcal{R}^{-1}\mathcal{P}\mathcal{M}^{-1}\mathbf{x}_{\rm{gal}}$ and then make use of the relation $\mathbf{x}_{\text{hel-ecl}}=(\cos \ell \cos b, \sin \ell \cos b, \sin b)$.

Carrying out this procedure and expanding to first order in $T$, we find
\begin{align}
(b_X,\lambda_X)&=(5\fdg536, 266\fdg840) + (0\fdg013,1\fdg397)\,T\\
(b_Y,\lambda_Y)&=(-59\fdg574,347\fdg340) + (0\fdg002 ,1\fdg375 )\,T\\
(b_Z,\lambda_Z)&=(-29\fdg811,180\fdg023) + (0\fdg001,1\fdg404)\,T.
\end{align}
As we will show in section~\ref{section:phenomenology}, the time dependence of $b_i$ and $\varpi$ can be ignored with little loss in accuracy.

\subsection{Comparisons}
\label{subsection:comparison}

It should be clear from the above derivations that the expressions for $\mathbf{u}_{\rm{E}}(t)$ in eqs.~\eqref{eq:uex1} and~\eqref{eq:uex2} are equivalent. We make the equivalence explicit by noticing that the components of $\epx$ and $\epy$ are related to $b_i$ and $\lambda_i$ through $\hat{\bepsilon}_{\mathbf{x}i} \equiv-\cos b_i \cos \lambda_i$ and $\hat{\bepsilon}_{\mathbf{y}i} \equiv-\cos b_i \sin \lambda_i$. A numerical resolution of these expressions confirms that this is true.

We now compare our expression with the two other expressions in the literature that are also valid to first order in $e$. Firstly, we compare with the expression of Lee, Lisanti and Safdi~\cite{Lee:2013xxa}. In order to compare expressions, we need to introduce new notation that matches theirs: they define $\lambda_p=\varpi-180\degree$, $g=\omega(t-t_p)$, where $t_p$ is the time at perihelion and define the time $t_1$ when the Earth is at the vernal equinox $\Upsilon$ $(\ell=0\degree)$. Their unit vectors are also different; they are related to ours by $\epone=\epy$, $\eptwo=-\epx$. To obtain their expression, we need to relate $t_p$ to $t_1$: inverting eq.~\eqref{eq:centre} gives $g=\ell-\lambda_p-\pi+2e\sin(\ell-\lambda_p)$. Then, $t_p$ is expressed in terms of $t_1$ by evaluating this expression at $t=t_1$, $\ell=0\degree$. From this we find (to first order in $e$) that $L=\omega (t-t_1)-2e\sin\lambda_p$. Substituting this into eq.~\eqref{eq:uex1} and expanding to first  order in $e$, we obtain
\begin{equation}
\begin{split}
\mathbf{u}_{\mathrm{E}}(t)&=\langle u_{\rm{E}}\rangle [\cos\omega(t-t_1)-2 e \sin \lambda_p \sin\omega(t-t_1) \\
&\qquad\qquad-e\cos\left(2 \omega (t-t_1)-\lambda_p\right)] \epone\\
&+\langle u_{\rm{E}}\rangle [\sin\omega(t-t_1)-2 e \sin \lambda_p \cos\omega(t-t_1)\\
&\qquad\qquad-e\sin\left(2 \omega (t-t_1)-\lambda_p\right)]\eptwo
\end{split}
\end{equation}
with $\langle u_{\rm{E}}\rangle=a \omega=29.79$~km/s. This expression matches the expression in section~B.1 of~\cite{Lee:2013xxa}. Furthermore, using $T=0.13$ in eqs.~\eqref{eq:epx} and~\eqref{eq:epy} gives the same numerical values for the unit vectors $\epone$ and $\eptwo$ in 2013 as quoted in~\cite{Lee:2013xxa}.

Secondly, we compare with the expression in appendix~B of Lewin and Smith~\cite{Lewin:1995rx}. This expression is commonly cited in the literature but it was suggested in~\cite{Lee:2013xxa} that it is incorrect. We demonstrate that this is the case. Lewin and Smith retain the full expression for $\ell$ in the velocity, rather than expanding to first order $e$ as in eq.~\eqref{eq:uex2}. Differentiating eq.~\eqref{eq:r-gal} and expanding only $r$, $\dot{r}$ and $\dot{\ell}$ to first order in $e$, we obtain
\begin{equation}
\mathbf{u}_{\mathrm{E}\,i}(t)=\langle u_{\rm{E}}\rangle\cos b_i\left[ \sin(\ell-\lambda_i)-e\cos(\lambda_i-\lambda_0)\right]\;,
\end{equation}
where following~\cite{Lewin:1995rx}, we have introduced the latitude of the orbit's minor axis: $\lambda_0=\varpi-270\degree\approx12\fdg9$. Comparing with Lewin and Smith's expression, we find that the numerical values of the latitude $b_i$ and longitude $\lambda_i$ are similar, differing by $\sim 0\fdg5$ (the sign of $b_i$ is opposite because we use a heliocentric system while they use a geocentric system). While the leading order term matches theirs, the first order in $e$ correction is different: we find that this correction is $-\cos(\lambda_i-\lambda_0)$ rather than $-\sin(\ell-\lambda_0)$, as stated in~\cite{Lewin:1995rx}.

\section{Phenomenological consequences}
\label{section:phenomenology}

We now consider the phenomenological consequences of using different expressions for $\mathbf{u}_{\rm{E}}(t)$ when calculating observable quantities at direct detection experiments. 

Firstly, we consider the change in the unmodulated event rate. Typically, the consequences of ignoring the $T$ and $e$ dependence in eqs.~\eqref{eq:uex1} and~\eqref{eq:uex2} are very small. For instance, considering the total rate $R$ at the LUX experiment~\cite{Akerib:2013tjd}, the percentage change in $R$ calculated with  $T$ and $e$ dependence compared to $R$ calculated without the $T$ and $e$ dependence is always less than $0.2\%$. This holds for the full dark matter mass range ($\text{few GeV}\lesssim m_{\rm{DM}} \lesssim \text{multi TeV}$), and for different interactions, including standard elastic scattering, momentum dependent scattering~\cite{Chang:2009yt} and inelastic scattering~\cite{TuckerSmith:2001hy} (see e.g.~\cite{Frandsen:2013cna} for further details of these interactions). In comparison, the change in $R$ when $v_0$ is varied by 30~km/s is $\sim50\%$ for elastic and momentum dependent scattering (with $m_{\chi}=10$~GeV) and $\sim80\%$ for inelastic dark matter (with $m_{\rm{DM}}=100$~GeV and $\delta=100$~keV). Therefore, when considering the unmodulated rate, we conclude that it is always a good approximation to ignore the $T$ and $e$ dependence in eqs.~\eqref{eq:uex1} and~\eqref{eq:uex2} because the change in $R$ owing to uncertainties in other astrophysical parameters is much greater. 

Secondly, we consider the consequences for the modulated event rate. For a differential cross-section $d \sigma /d E_{\rm{R}}\propto v^{-2}$, which is the form assumed by experiments when they set a spin-independent or spin-dependent limit, the Earth's velocity affects the rate $R$ through the velocity integral
\begin{equation}
g(v_{\min},t)=\int_{v_{\rm{min}}}d^3v\, \frac{f_{\rm{gal}}(\mathbf{v}+\mathbf{v}_{\rm{Earth}}(t))}{v}\;,
\end{equation}
where (for elastic scattering)
\begin{equation}
v_{\rm{min}}=\frac{m_{\rm{N}}+m_{\rm{DM}}}{m_{\rm{DM}}}\sqrt{\frac{E_{\rm{R}}}{2 m_{\rm{N}}}}\;.
\end{equation}
We assume the `Standard Halo Model' in which $f_{\rm{gal}}(v)$ is simply an exponential with velocity dispersion $\sqrt{3/2}v_0$ and a hard cut-off at the galactic escape speed (taken as 533~km/s~\cite{Piffl:2013mla}). In this case, $g(v_{\min},t)$ has a simple analytic form~\cite{Savage:2008er, McCabe:2010zh}. We also assume that all of the time dependence in $R$ arises from $\mathbf{u}_{\rm{E}}(t)$, ignoring other sources of time dependence such as gravitational focussing~\cite{Lee:2013wza}.

\begingroup
\begin{table}[t!]
\begin{ruledtabular}
\begin{tabular}{c c c c}
 &$\Delta t^{\text{1 yr}}_{\text{peak}}$ & $\Delta t^{\text{14 yrs}}_{\text{peak}}$&$\Delta A$  \\[0.5ex] \hline
eq.~\eqref{eq:uex1} &0.06 & 0.05 &-0.02\% \\ 
eq.~\eqref{eq:uex2}  &0.06 & 0.05 &-0.02\%\\ 
$e=0$& -1.17 & -1.18 & -0.39\% \\
$T=0$& 0.06&0.26 &  -0.02\%\\ 
$e=T=0$ & -1.17&-0.97 & -0.39\% \\ 
$\Delta v_0=30$~km/s& -0.78& -0.77 & 19.75\%\\  
$\Delta \mathbf{v}_{\rm{pec}}=(1.2,2.0,0.6)$~km/s& 0.63& 0.63 & -0.87\%\\  
eq. of Lewin \& Smith& 1.28& 1.49 &-0.01\% \\  
\end{tabular}
\caption{The second and third columns show the difference (in days) in the peak day $\Delta t_{\rm{peak}}$ (after 1 year and 14 years from J2000.0 respectively), and the fourth column shows the percentage change in the amplitude $A$ between the exact expression for $\mathbf{u}_{\rm{E}}(t)$ and the expressions and approximations listed in the first column. The first two rows show that eqs.~\eqref{eq:uex1} and~\eqref{eq:uex2} are a very good approximation. The dominant change in the peak day arises when the $e$ dependence is ignored. Varying $v_0$ has the largest effect on the amplitude. Lewin \& Smith's expression~\cite{Lewin:1995rx} predicts a peak day that is incorrect by 1.5 days.}
\label{tab:pheno}
\end{ruledtabular}
\end{table}
     \endgroup

Table~\ref{tab:pheno} lists the difference in the peak day and the amplitude of the modulated rate between the exact expression for $\mathbf{u}_{\rm{E}}(t)$ and the expressions and approximations listed in the first column. By `exact' expression, we mean that eq.~\eqref{eq:r-ecl} is differentiated numerically with all time dependence retained and $r$ and $\ell$ are expanded to third order in $e$. The second and third column show the difference (in days) in the peak day \mbox{$\Delta t_{\rm{peak}}=t^{\rm{exact}}_{\rm{peak}}-t^{\rm{approx}}_{\rm{peak}}$} after 1~year and 14~years from J2000.0 respectively. In 2014, the exact expression predicts that the modulation peaks at about 7.45pm GMT on 1st June.  The fourth column shows the percentage change in the amplitude \mbox{$\Delta A=100(A_{\rm{exact}}-A_{\rm{approx}})/A_{\rm{exact}}$} assuming that $E_{\rm{R}}=3$~keV (the threshold of LUX~\cite{Akerib:2013tjd} and XENON100~\cite{Aprile:2012nq}) and $m_{\rm{N}}$ is the mass of a xenon nucleus. We calculate $\Delta A$ for $m_{\rm{DM}}=10,100$ and~1000~GeV, quoting the largest value in Table~\ref{tab:pheno}. The values for the other choices of $m_{\rm{DM}}$ typically differ by a factor of a few.

The first two rows of table~\ref{tab:pheno} show that the expansion of $\mathbf{u}_{\rm{E}}(t)$ to first order in $e$ and $T$ in eqs.~\eqref{eq:uex1} and~\eqref{eq:uex2} is a very good approximation. In these eqs., we have ignored the $n$ dependence of $\varpi$ and the $T$ dependence of $b_i$ in eq.~\eqref{eq:uex2} as they are subdominant. In the next three rows, we consider the effect of ignoring the $e$ and $T$ dependence in eqs.~\eqref{eq:uex1} and~\eqref{eq:uex2}. The modulation peaks over a day earlier when $e=0$ but the change in the modulation amplitude remains relatively small. Ignoring the $T$ dependence leads to a smaller change but over a timescale of decades, the difference in the peak day can become comparable to the effect of ignoring $e$. The change in the amplitude remains small. When both terms are ignored, the difference from the exact expression is the linear sum of the individual differences. In the sixth and seventh rows, we show the changes from varying $v_0$ and $\mathbf{v}_{\rm{pec}}$ within their experimental error. The change in the peak day is slightly smaller than the effect of ignoring the first order $e$ correction while the change in the amplitude is significantly larger. In the final row, we compare with the expression of Lewin and Smith. We find that their expression predicts a peak day which is incorrect by 1.5 days (in 2014).

In conclusion, we have demonstrated that expanding $\mathbf{u}_{\rm{E}}(t)$ to first order in $e$ and $T$ leads to results that are very close to the results from the exact expression. When interested in predictions of the peak day of the modulation, we conclude that the $e$ dependence should be retained because its effect is larger than the effect from other approximations or varying other astrophysical parameters. If interested in changes of the amplitude, the $v_0$ dependence is by far the most dominant. The $T$ dependence becomes increasingly important over a timescale of decades from J2000.0.

These results agree with other phenomenological studies in the literature. Green~\cite{Green:2003yh} found that the peak day of the modulation signal can change by up to 10~days if an overly simplistic form of $\mathbf{u}_{\rm{E}}(t)$ is used. Meanwhile, Lee, Lisanti and Safdi~\cite{Lee:2013xxa} demonstrated that an expansion of $\mathbf{u}_{\rm{E}}(t)$ to first order in $e$ is required to accurately calculate the amplitude of the leading modes beyond annual modulation.

\section{Conclusions}
The Earth's velocity relative to the Sun expressed in galactic coordinates, $\mathbf{u}_{\rm{E}}(t)$, is crucial in determining the scattering rate at direct detection experiments. Somewhat surprisingly, many non-equivalent expressions for $\mathbf{u}_{\rm{E}}(t)$ are used in the literature. In sections~\ref{section:orbit} and~\ref{section:Eu} we have provided a rigorous derivation of two separate (but equivalent) forms: in eq.~\eqref{eq:uex1} and eq.~\eqref{eq:uex2} respectively. These expressions are series expansions to first order in the eccentricity $e$ and epoch of date $T$. The dependence on $T$ arises from the precession of the equinoxes and has not previously been discussed in the context of $\mathbf{u}_{\rm{E}}(t)$. 

In section~\ref{section:Eu} we compared our result with the two other expressions in the literature that are valid to first order in $e$. We found that our expressions are equivalent to the result of Lee, Lisanti and Safdi~\cite{Lee:2013xxa} (after a suitable change of notation) but disagree with the result of Lewin and Smith~\cite{Lewin:1995rx}. The incorrect equation of Lewin and Smith predicts that the peak day of the modulated rate differs by 1.5 days from that calculated with eq.~\eqref{eq:uex1} (or equivalently, eq.~\eqref{eq:uex2}). The result of Lewin and Smith has been used as the {\it de facto} standard (see e.g.~\cite{Green:2003yh}). 

We considered the phenomenological consequences of using different expressions for $\mathbf{u}_{\rm{E}}(t)$ in section~\ref{section:phenomenology}. For the unmodulated rate, we found that varying astrophysical parameters (such as the Sun's circular speed $v_0$) within their uncertainty leads to a much larger change in the calculated rate, compared to the modest change when using the tree-level expression for $\mathbf{u}_{\rm{E}}(t)$, which ignores $e$ and $T$ dependence. In comparison, the $e$ dependence should be retained for modulation studies that require an accurate prediction of the peak day of the modulated rate. Ignoring the $e$ dependence leads to a larger change than varying astrophysical parameters within their uncertainty (see table~\ref{tab:pheno}). The $T$ dependence becomes important over a timescale of decades.

\section*{Acknowledgements}
\noindent CM thanks C\'eline B\oe hm, Jonathan Davis, Anne Green and Felix Kahlhoefer for valuable discussions. This work has been partially supported by the European Union FP7 ITN INVISIBLES (Marie Curie Actions, PITN- GA-2011- 289442).

\appendix

\section{Calculating the fractional day number}
\label{sec:epoch}
For the zero of our reference time, we take the J2000.0 epoch, which is 12h Terrestrial time (TT) on 1st January 2000. Universal time (UT) or Greenwich Mean Time (GMT) can be used with negligible error. At J2000.0, the fractional day number $n=0$.

For a calendar date with year $Y$, month $M$ (1 for January, 2 for February, etc.) and (fractional) day of the month $D$ relative to midnight GMT, the fractional day is~\cite{Meeus2000}
\begin{equation}
n=\lfloor 365.25 \tilde{Y}\rfloor+\lfloor30.61 (\tilde{M}+1)\rfloor+D-730563.5\;,
\end{equation}
where the floor function $\lfloor x \rfloor$ gives the largest integer less than or equal to $x$ and
\begin{align}
\tilde{Y} &=
  \begin{cases}
   Y-1 & \text{if } M=1\text{ or }2 \\
   Y       & \text{if } M >2
  \end{cases}\\
  \tilde{M} &=
  \begin{cases}
   M+12 & \text{if } M=1\text{ or }2 \\
   M       & \text{if } M >2
  \end{cases}\;.
\end{align}
For example, for 6~p.m.\ GMT on 31st January 2009, $\tilde{Y}=2008$, $\tilde{M}=13$ and $D=31.75$ resulting in $n=3318.25$. 

\section{Rotation matrices}
\label{app:rotation}
In this appendix we give the elements of the rotation matrices used in section~\ref{section:Eu}. The axes shown in fig.~\ref{fig:ellipse} are the $x$ and $y$-axes of the heliocentric ecliptic rectangular coordinate system. Geocentric ecliptic rectangular coordinates are related to these coordinates through~\cite{AA2014}
\begin{equation}
\mathbf{x}_{\text{geo-ecl.}}=-\mathbf{x}_{\text{hel-ecl.}}\;.
\end{equation}

Equatorial rectangular coordinates are related to geocentric ecliptic rectangular coordinates through the relation \mbox{$\mathbf{x}_{\rm{equat.}}=\mathcal{R}\,\mathbf{x}_{\text{geo-ecl.}}$}, where~\cite{AA2014}
\begin{equation}
\mathcal{R}=
 \begin{pmatrix}
1 &0 & 0 \\
0 & \cos \epsilon  &-\sin\epsilon \\
 0 & \sin \epsilon  & \cos \epsilon
 \end{pmatrix}\;.
\end{equation}
Here $\epsilon=23\fdg4393-0.0130 \,T$ is the obliquity of the ecliptic and $T$ is the epoch of date.

Galactic rectangular coordinates are related to equatorial rectangular coordinates at J2000.0 through the relation \mbox{$\mathbf{x}_{\rm{gal.}}=\mathcal{M}\,\mathbf{x}_{\rm{equat.}}(\text{J}2000.0)$}, where the elements of $\mathcal{M}$ are
\begin{equation}
\begin{split}
\mathcal{M}_{11}&= -\sln \san-\cln\can\sdn \\
\mathcal{M}_{12}&= \sln \can-\cln \san \sdn \\
\mathcal{M}_{13}&=  \cln \cdn\\
\mathcal{M}_{21}&= \cln \san-\sln \can\sdn   \\
\mathcal{M}_{22}&=  -\cln\can-\sln\san\sdn\\
\mathcal{M}_{23}&=  \sln\cdn\\
\mathcal{M}_{31}&=   \can\cdn   \\
\mathcal{M}_{32}&=  \san\cdn  \\
\mathcal{M}_{33}&= \sdn\;.
\end{split}
\end{equation}
Here $\alpha_{\rm{GP}}$ and $\delta_{\rm{GP}}$ are the right ascension and the declination of the north galactic pole. By international convention, the J2000.0 right ascension and declination are \mbox{$(\alpha_{\rm{GP}},\delta_{\rm{GP}})=(192\fdg85948,27\fdg12825)$}. Also required is the longitude of the north celestial pole, which in J2000.0 galactic coordinates, is $l_{\rm{CP}}=122\fdg932$~\cite{Binney1998}.

The rotation matrix to rotate equatorial rectangular coordinates from J2000.0 to epoch of date $T$ is \mbox{$x_{\rm{equat.}}(T)=\mathcal{P}\,x_{\rm{equat.}}(\text{J}2000.0)$}, where the elements are
\begin{equation}
\begin{split}
\mathcal{P}_{11}&= \cze\ct\cz-\sze\sz \\
\mathcal{P}_{12}&= -\sze\ct\cz-\cze\sz \\
\mathcal{P}_{13}&=-\st\cz \\
\mathcal{P}_{21}&=  \cze\ct\sz+\sze\cz  \\
\mathcal{P}_{22}&= -\sze\ct\sz+\cze\cz\\
\mathcal{P}_{23}&=-\st\sz \\
\mathcal{P}_{31}&=  \cze\st   \\
\mathcal{P}_{32}&= -\sze\st \\
\mathcal{P}_{33}&=\ct\;.
\end{split}
\end{equation}
Here
\begin{align}
\zeta_{\rm{A}}&=2306\farcs083227\,T+0\farcs298850\,T^2\\
z_{\rm{A}}&=2306\farcs077181\,T+1\farcs092735\,T^2\\
\zeta_{\rm{A}}&=2004\farcs191903\,T-0\farcs429493\,T^2
\end{align}
are the traditional equatorial precession angles~\cite{AA2014}. 

\bibliography{ref}
\bibliographystyle{ArXiv}

\end{document}